\documentclass[aps,pra,twocolumn,superscriptaddress]{revtex4-1}

\usepackage[shortlabels]{enumitem}
\usepackage{graphicx}
\usepackage{epsfig}
\usepackage{comment}
\usepackage{amsmath}
\usepackage{amsfonts}
\usepackage{amssymb}
\usepackage{bm}
\usepackage{mathtools}

\usepackage{ytableau}

\UseRawInputEncoding 

\usepackage[T1,T2A]{fontenc}
\usepackage[utf8]{inputenc}
\usepackage[russian,english]{babel}

\usepackage{hyperref}
\usepackage{dsfont}
\usepackage{lipsum}

\usepackage{tikz}
\usepackage{circuitikz}
\usetikzlibrary{arrows, shapes}
\usepackage{pgfplots}
\pgfplotsset{compat=1.14}

\hypersetup{%
    colorlinks=true, linktocpage=true, pdfstartpage=1, pdfstartview=FitV
}
\usepackage[capitalise]{cleveref}

\newcommand{\ket}[1]{| #1 \rangle}

\DeclareMathOperator\Tr{Tr}

\begin{document} 
\title{Symmetry shapes thermodynamics of macroscopic quantum systems}
\author{Vasco Cavina}
\email{vasco.cavina@sns.it}
\affiliation{Department of Physics and Materials Science, University of Luxembourg, L-1511 Luxembourg, Luxembourg}
\affiliation{NEST, Scuola Normale Superiore and Istituto Nanoscienze-CNR, I-56126 Pisa, Italy}
\author{Ariane Soret}
\email{ariane.soret@gmail.com}
\affiliation{Department of Physics and Materials Science, University of Luxembourg, L-1511 Luxembourg, Luxembourg}
\author{Timur Aslyamov}
\affiliation{Department of Physics and Materials Science, University of Luxembourg, L-1511 Luxembourg, Luxembourg}
\author{Krzysztof Ptaszy\'{n}ski}
\affiliation{Department of Physics and Materials Science, University of Luxembourg, L-1511 Luxembourg, Luxembourg}
	\affiliation{Institute of Molecular Physics, Polish Academy of Sciences, Mariana Smoluchowskiego 17, 60-179 Pozna\'{n}, Poland}
\author{Massimiliano Esposito}
\affiliation{Department of Physics and Materials Science, University of Luxembourg, L-1511 Luxembourg, Luxembourg}

\begin{abstract}
We derive a systematic approach to the thermodynamics of quantum systems based on the underlying symmetry groups.
We show that the entropy of a system can be described in terms of group-theoretical quantities that are largely independent of the details of its density matrix.
We apply our technique to generic $N$ identical interacting $d$-level quantum systems. Using permutation invariance, we find that, for large $N$, entropy displays a universal large deviation behavior with a rate function $s(\boldsymbol{x})$ that is completely independent of the microscopic details of the model, but depends only on the size of the irreducible representations of the permutation group $\text{S}_N$.
In turn, the partition function is shown to satisfy a large deviation principle with a free energy $f(\boldsymbol{x})=e(\boldsymbol{x})-\beta^{-1}s(\boldsymbol{x})$, where $e(\boldsymbol{x})$ is a rate function that only depends on the ground state energy of particular subspaces determined by group representation theory.
We apply our theory to the transverse-field Curie-Weiss model, a minimal model of phase transition exhibiting an interplay of thermal and quantum fluctuations.
\end{abstract}
\maketitle
Symmetries play a fundamental role in shaping physical theories, from quantum mechanics to thermodynamics \cite{landau2013course}.
In the context of nonequilibrium physics, symmetries are crucial for the understanding of stochastic energetics
\cite{polettini2016conservation,rao2018conservation}, dynamical phase transitions \cite{heyl2018dynamical, heyl2014dynamical,knap2017} and quantum thermodynamics \cite{aron2018non, fei2019group, cavina2023convenient}.
Studying the entropic, energetic, or dynamic signatures of underlying symmetries in quantum systems is an active field of research, from fundamental questions about entropy scalings, ground state properties, and thermalization \cite{gour2018, shpielberg2022, roos2019, decamp2017, ueda2020, aragoneses2023,yu2024}, to the optimization of quantum computing or numerical simulation procedures \cite{esposito2005emergence, esposito2005jstat, franco2018, nori2018, mansky2023permutationinvariant}. This research is gaining momentum due to rapid experimental advances \cite{ueda2020}, particularly in cold atoms \cite{dlaska2022, islam2022}.

In this letter, we give a systematic description of how the underlying symmetry of a quantum system determines its entropic and energetic profile.
First, we show that if the density matrix of a system is invariant under the action of a group $G$, its von Neumann entropy simplifies into a contribution that only depends on a structural property of the group, namely the dimension of its irreducible representations, and another that is model dependent but can be bounded from above by other group theoretical quantities.
Second, 
we consider many-body quantum systems composed of $N$ identical interacting $d$-level systems and invariant under the action of the permutation group $G=\text{S}_N$. 
At equilibrium, we prove that the only extensive contributions to entropy are completely independent of the microscopic details of the system and are only shaped by the structure of $\text{S}_N$.
The extensive contributions to energy are, in turn, given by the ground states of particular restrictions of the Hamiltonian,  defined from group representation theory.
Third, we illustrate the usefulness of our results on the transverse-field Curie-Weiss model, which -- despite the apparent simplicity of its Hamiltonian -- exhibits a highly nontrivial behavior resulting from the interplay of quantum and thermal fluctuations~\cite{chayes2008phase, jorg2010energy}.

\medskip

{\it Von Neuman entropy using group theory.---}If a quantum model is invariant under the action of a given finite group $G$ 
\footnote{Formally, when we say that a group $G$ acts on the Hilbert space of the system $\mathcal{H}$, we mean that there is a group homomorphism between $G$ and $ GL(\mathcal{H})$, the group of linear invertible operators acting on $\mathcal{H}$. To simplify the notation, we will never explicitly write this homomorphism and just treat $G$ as a subgroup of $ GL(\mathcal{H})$. With this in mind, the elements of $G$ will be accompanied by a hat to remark that they are operators acting on $\mathcal{H}$.}, 
we can exploit representation theory to constrain the structure of the density matrix. This will be instrumental in understanding how symmetry-related effects influence the thermodynamics of the system.
 Maschke's theorem \cite{fultonharris,sagan} allows us to decompose the Hilbert space $\mathcal{H}$ into a direct sum of {\it representations} of $G$,
that is, subspaces of $\mathcal{H}$ that remain invariant under the action of any element $\hat{g} \in G$.
A representation that cannot be further decomposed (i.e., does not contain invariant subspaces on its own) is called {\it irreducible}.
The Hilbert space has a unique decomposition in terms of irreducible representations, that are hereby called $ V_{k}^{\boldsymbol{\lambda}}$
:
\begin{equation} \label{eq:decomp0}
\mathcal{H} = \bigoplus_{\boldsymbol{\lambda}}  \bigoplus_{k=1}^{\deg \boldsymbol{\lambda}} V_{k}^{\boldsymbol{\lambda}}\,,
\end{equation}
where $\boldsymbol{\lambda}$ is a suitable index \footnote{The domain of $\boldsymbol{\lambda}$ depends on the group of interest. In the case of the group $\text{S}_N$ it takes the form of a vector, as we will see later.} labeling the irreducibles of $G$
in the decomposition of $\mathcal{H}$. The same irreducible can appear with multiplicity denoted $\deg{\boldsymbol{\lambda}}$ in Eq. \eqref{eq:decomp0}.
A state $\hat{\rho}$ is invariant under the action of $G$ if $[\hat{g}, \hat{\rho}]=0$ for all $\hat{g} \in G$ \footnote{See footnote [24].}, so it can be fully characterized by Schur's lemma \cite{fultonharris,sagan}.
The latter implies that the density matrix is block diagonal in the sectors with fixed~$\boldsymbol{\lambda}$.
Defining the total probability associated with a given block as
$p_{\boldsymbol{\lambda}} = \Tr[\hat{\rho} \sum_k \hat{\Pi}_{\boldsymbol{\lambda},k}]$, with
$\hat{\Pi}_{\boldsymbol{\lambda}, k}$ the projector on $V_{k}^{\boldsymbol{\lambda}}$, we obtain
\begin{equation} \label{eq:decomp1}
\hat{\rho} = \sum_{\boldsymbol{\lambda}} p_{\boldsymbol{\lambda}} \hat{\rho}_{\boldsymbol{\lambda}}\,,
\end{equation}
where 
$\hat{\rho}_{\boldsymbol{\lambda}}\equiv\sum_{k,k'} \hat{\Pi}_{\boldsymbol{\lambda},k} \hat{\rho}  \hat{\Pi}_{\boldsymbol{\lambda},k'}/p_{\boldsymbol{\lambda}}$ are the conditional (normalized) density matrices associated to the blocks. 
A second implication of Schur's lemma is that every subblock of $\hat{\rho}_{\boldsymbol{\lambda}}$ with fixed $k,k'$ is proportional to the identity matrix,
\begin{equation} \label{eq:deftilted}
\hat{\Pi}_{\boldsymbol{\lambda},k} \hat{\rho}_{\boldsymbol{\lambda}} \hat{\Pi}_{\boldsymbol{\lambda},k'}  = \frac{1}{\dim \boldsymbol{\lambda}}[\tilde{\rho}_{\boldsymbol{\lambda}}]_{kk'} \mathds{I}_{\boldsymbol{\lambda}kk'},
\end{equation}
where we gathered the proportionality coefficients in a new matrix $\tilde{\rho}_{\boldsymbol{\lambda}}$  defined on a reduced Hilbert space spanned by the indices $k,k' \in [1, \deg \boldsymbol{\lambda}]$, and where  $\mathds{I}_{\boldsymbol{\lambda}kk'}$ is the identity map between $V_{\boldsymbol{\lambda},k} $ and $ V_{\boldsymbol{\lambda},k'}$.
The normalization $\dim \boldsymbol{\lambda} = \dim V_{\boldsymbol{\lambda},k} = \dim V_{\boldsymbol{\lambda},k'}$ ensures that $\Tr[\tilde{\rho}_{\boldsymbol{\lambda}}]=1$.
Using the new ``coarse-grained'' state $\tilde{\rho}_{\boldsymbol{\lambda}} $, we may rewrite $\hat{\rho}_{\boldsymbol{\lambda}} $ as
\begin{equation} \label{eq:decomp2}
\hat{\rho}_{\boldsymbol{\lambda}} = \frac{1}{\dim \boldsymbol{\lambda} } \tilde{\rho}_{\boldsymbol{\lambda}}  
\otimes \mathds{I}_{\boldsymbol{\lambda}}\,.
\end{equation}

We now use Eqs. \eqref{eq:decomp1} and \eqref{eq:decomp2} to compute the von Neumann entropy $S(\hat{\rho}) \equiv - \Tr[\hat{\rho} \ln \hat{\rho} ]$. 
Since the blocks with different $\boldsymbol{\lambda}$ are disconnected (see Eq. \eqref{eq:decomp1}), the final result will be a sum over the contributions of the different blocks. In addition, the inside-block degeneracy emerging from Eq. \eqref{eq:decomp2} yields an additional $\dim \boldsymbol{\lambda}$ factor to the entropy, resulting in
(see Appendix \ref{app:entropy})
\begin{equation} \label{eq:entropy}
S(\hat{\rho}) = \sum_{\boldsymbol{\lambda}} p_{\boldsymbol{\lambda}} 
\big[ \ln \dim \boldsymbol{\lambda} 
+ S(\tilde{\rho}_{\boldsymbol{\lambda}} ) \big] + H(p)\,,
\end{equation}
where $H(p) = -\sum_{\boldsymbol{\lambda}} p_{\boldsymbol{\lambda}} \ln p_{\boldsymbol{\lambda}}$ is the Shannon entropy of the classical probability distribution $p = \{p_{\boldsymbol{\lambda}}\}_{\boldsymbol{\lambda}}$.
The first contribution in Eq. \eqref{eq:entropy} depends on the dimension of a given irreducible representation, i.e. a universal property of the group $G$.
On the contrary, the second and third terms depend on the details of $\hat{\rho}$, yet they can be bounded from above by other state-independent parameters coming from Eq. \eqref{eq:decomp0}.
In fact, the von Neumann entropy is maximal on the totally mixed state $\mathcal{E}_{\boldsymbol{\lambda}}$, that is $S(\tilde{\rho}_{\boldsymbol{\lambda}}) \leq S(\mathcal{E}_{\boldsymbol{\lambda}})$. Since the coarse-grained space with fixed $\boldsymbol{\lambda}$ has dimension $\deg \boldsymbol{\lambda}$ (see below Eq. \eqref{eq:deftilted}), we obtain
\begin{equation} \label{eq:bound1}
S(\tilde{\rho}_{\boldsymbol{\lambda}}) \leq \ln \deg \boldsymbol{\lambda}\,.
\end{equation}

Similarly, the Shannon entropy $H(p)$ can never exceed the value of the logarithm of the total number of irreducible representations appearing in the direct sum \eqref{eq:decomp0}, that we call $\Lambda$ 
\begin{equation} \label{eq:bound2}
H(p) \leq \ln \Lambda\,.
\end{equation}
Eqs. \eqref{eq:entropy}, \eqref{eq:bound1} and \eqref{eq:bound2} allow us to give a universal description of the von Neumann entropy of any quantum system given its symmetry group $G$.

{\it Universality of macroscopic entropy.---}%
For a collection of $N$ identical $d$-level systems, the Hilbert space has the form
\begin{equation} \label{eq:hilbert}
\mathcal{H}_N^{(d)} = \bigotimes_{i=1}^N \mathds{C}^{(i)}_d,
\end{equation}
where $\mathds{C}^{(i)}_d$ is the $d$-dimensional complex vector space of the $i$-th system. The set $G$ of symmetries of a given system depends on the properties of the system itself. Common examples in many-body physics are the cyclic symmetry (e.g. in Ising chains with periodic boundary conditions) and the permutation symmetry, associated, respectively, with the cyclic group $\text{Z}_N$ and to the symmetric group $\text{S}_N$.

From now on, we focus on the permutation group. Each element $\pi \in \text{S}_N$ acts in a simple way on the tensor product basis of $\mathcal{H}_N^{(d)}$, by exchanging its local components,
\begin{equation} 
    \pi \ket{e_1} \otimes \ket{e_2} ..\otimes \ket{e_N} = 
    \ket{e_{\pi(1)}} \otimes \ket{e_{\pi(2)}} ..\otimes \ket{e_{\pi(N)}} \, ,\label{eq:perm-states}
\end{equation}
where $\pi(j)$ is the particle occupying the position $j$ after the permutation took place and $\ket{e_{i}}$ with $e_i =1,... \:d$ is a basis vector for $\mathds{C}^{(i)}_d$.
To obtain a decomposition of the form \eqref{eq:decomp0} for $\mathcal{H}_N^{(d)}$ and $G = \text{S}_N$, let us first introduce the {\it occupation numbers} $\mu_j$, which represent the number of particles in the state $j$ for $j = 1, ... \: d$. We then introduce the subspaces $M_{\boldsymbol{\mu}}$ with fixed occupation number vector $\boldsymbol{\mu}=(\mu_1,...,\mu_d)$. By construction, the basis elements of $M_{\boldsymbol{\mu}}$ are vectors $\ket{e_1} \otimes \ket{e_2} ..\otimes \ket{e_N}$ such that the value $j$ is taken exactly $\mu_j$ times by the $e_i$'s, for $j=1,...,d$. From the relation \eqref{eq:perm-states}, it is clear that the subspaces $M_{\boldsymbol{\mu}}$ are invariant under the action of $\text{S}_N$, so we can preliminarily decompose
\begin{equation} \label{eq:module1}
\mathcal{H}_N^{(d)} = \bigoplus_{\boldsymbol{\mu}} M_{\boldsymbol{\mu}}\,,
\end{equation}
where $\pi M_{\boldsymbol{\mu}} \subseteq M_{\boldsymbol{\mu}}$ for all $\pi \in \text{S}_N$.
To further reduce the susbspaces $M_{\boldsymbol{\mu}}$, we do the crucial observation (see Appendix \ref{app:modules}) that they coincide with standard objects of representation theory, called {\it permutation modules} of index $\boldsymbol{\mu}$.
Permutation modules can be completely reduced, using textbook approaches \cite{sagan}, as
\begin{equation} \label{eq:module2}
M_{\boldsymbol{\mu}} = 
\bigoplus_{\boldsymbol{\lambda}} \bigoplus_{h=1}^{K_{\boldsymbol{\lambda}, \boldsymbol{\mu}}} \mathcal{S}^{\boldsymbol{\lambda}}_{h,\boldsymbol{\mu}}\,,
\end{equation}
where  $\mathcal{S}^{\boldsymbol{\lambda}}_{h, \boldsymbol{\mu}}$ are the irreducible representations of $\text{S}_N$.
While the superscript $\boldsymbol{\lambda}$ spans different irreducibles, the subscript $ \boldsymbol{\mu}$ keeps track of the original $\boldsymbol{\mu}$-sector and the subscript $h$ is a degeneracy index running between $1$ and a value $K_{\boldsymbol{\lambda},\boldsymbol{\mu}}$ that can be computed analytically and is known as the {\it Kostka number} \cite{sagan}.
The vector indices $\boldsymbol{\lambda}, \boldsymbol{\mu}$ are $d$-dimensional partitions of $N$ (decreasing vectors of positive integers summing to $N$, $\sum_{j=1}^d \mu_j = \sum_{j=1}^d \lambda_j = N$), and, together with $h$ in Eq. \eqref{eq:module2}, constitute a complete set of quantum numbers labeling the irreducibles in the decomposition of $\mathcal{H}_N^{(d)} $.

To study the behavior of the von Neumann entropy of a permutation-invariant state in the large $N$ limit we apply Eqs. \eqref{eq:entropy}, \eqref{eq:bound1} and \eqref{eq:bound2} to the decomposition \eqref{eq:module1}. 
After replacing $k$ with the specific notation for $\text{S}_N$ given by $\boldsymbol{\mu},h$, the asymptotic of the entropic contributions can be obtained from known combinatorial formulae: 
\begin{itemize}
    \item $\dim \boldsymbol{\lambda}$ is given by the {\it Hook length formula} \cite{fultonharris}, a combinatorial rule based on the concept of {\it Young tableau}. The Vershik-Kerov theorem and related results \cite{kerov1986, bufetov-2,Nazarov2023} show that $\dim \boldsymbol{\lambda}$ is exponentially large in $N$ for $N \rightarrow \infty$.
    %
    After introducing a rescaled variable $\boldsymbol{x} \equiv \frac{\boldsymbol{\lambda}}{N}$ and an intensive volume entropy associated with $\dim \boldsymbol{\lambda}$, we find (see Appendix \ref{app:semistand})
    \begin{equation} \label{eq:rateent}
    s(\boldsymbol{x}) \equiv  \lim_{N \to \infty} \frac{1}{N} \ln \dim (N \boldsymbol{x})  = -  \sum_{j=1}^d x_j \ln x_j\,, \hspace{-0.52cm}
    \end{equation}
    where $s(\boldsymbol{x})$ turns out to be a Shannon entropy over the rescaled variable $\boldsymbol{x}$. 
  \item  $\deg \boldsymbol{\lambda}$ is the multiplicity of the irreducible labeled by the partition $\boldsymbol{\lambda}$, which can be obtained from Eqs. \eqref{eq:module1}, $\eqref{eq:module2}$. An upper bound for this quantity is derived (see Appendix \ref{app:semistand}) by noting that it coincides with the logarithm of the Schur polynomial of index $\boldsymbol{\lambda} $ and all arguments equal to $1$ \cite{fultonharris}
    \begin{equation} 
    \label{eq:limit1} \ln \deg \boldsymbol{\lambda} = \ln \sum_{\boldsymbol{\mu}} K_{\boldsymbol{\lambda}, \boldsymbol{\mu}} \leq  \frac{d (d-1)}{2} \ln N\,,
    \end{equation}
    so the contribution of $\deg \boldsymbol{\lambda}$ is non-extensive, as long as the thermodynamic limit does not involve any scaling of the local Hilbert spaces dimension $d$.
    \item  $\Lambda$ is the number of distinct irreducible representations appearing in Eq. \eqref{eq:module2}.
    If the Hilbert spaces of the single constituents have dimension $d$, then  $\boldsymbol{\lambda}$ spans all the partitions of $N$ composed by $d$ elements. This quantity has a known asymptotic value for large $N$  \cite{knessl1990partition}, that we can use to derive
    \begin{equation} \label{eq:limit3}
        \ln \Lambda  \sim (d-1) \ln N \, .
    \end{equation}
    
\end{itemize}
Plugging equations \eqref{eq:limit1}, \eqref{eq:limit3} into Eqs. \eqref{eq:entropy}, \eqref{eq:bound1}, \eqref{eq:bound2} it is clear that the only extensive contribution to $S(\hat{\rho})$ comes from the term proportional to $\dim \boldsymbol{\lambda}$, thus
\begin{equation} \label{VNentropy}
\lim_{N \to \infty} \frac{S(\hat{\rho})}{N} = \lim_{N \to \infty} \sum_{\boldsymbol{\lambda}} \frac{p_{\boldsymbol{\lambda}}}{N} \ln \dim {\boldsymbol{\lambda}} =  \int  p(\boldsymbol{x}) s(\boldsymbol{x}) d\boldsymbol{x}\,,
\end{equation}
where $p(\boldsymbol{x}) \equiv N p_{\boldsymbol{\lambda}= N \boldsymbol{x}}$.
We stress that $s(\boldsymbol{x})$ is a universal property of $\text{S}_N$, and as such is completely independent from any detail of the state $\hat{\rho}$, that are gauged only through $p(\boldsymbol{x})$.

To show the versatility of our approach, we complete this section with a concise discussion of the cyclic group case $G = \text{Z}_N$. 
Since $\text{Z}_N$ is an abelian group, the dimension of irreducible representations $\dim \boldsymbol{\lambda}$ is always $1$. 
The number of distinct irreducibles $\Lambda$  can easily be shown to be bounded from above by the number of elements of $\text{Z}_N$ \footnote{The number of distinct irreducible representations of a group corresponds to the number of {\it conjugacy classes} of that group \cite{fultonharris}. For $\text{Z}_N$ and for abelian groups in general this corresponds to the number of elements of the group.
}, that is equal to $N$.
We conclude that the contributions related to $\ln \text{dim}\boldsymbol{\lambda}$ and $H(p)$ are sub-extensive for $\text{Z}_N$, meaning that the only extensive contributions to the entropy come from $S(\tilde{\rho}_{\boldsymbol{\lambda}})$ and are strongly state-dependent, as opposed to what happens for the permutation group $\text{S}_N$.

{\it Macroscopic equilibrium thermodynamics.---}A system at equilibrium is described by the free energy
\begin{equation} \label{eq:partition}
   F \equiv - \frac{1}{\beta} \ln \Tr[e^{-\beta \hat{H}}]\,,
\end{equation}
where $\hat{H}$ is the Hamiltonian and $\beta$ is the inverse temperature of the system (we set $k_B=1$).
If the Hamiltonian is invariant under the action of a generic group $G$, i.e.
$[\hat{H}, \hat{g}] = 0$ for all $\hat{g} \in G$ we can characterize $\hat{H}$ using Schur's lemma.
The procedure for decomposing $\hat{H}$ is even easier than the one followed for states in 
Eqs. \eqref{eq:decomp1}, \eqref{eq:deftilted}, \eqref{eq:decomp2}.
In the latter case, we needed to interpret $\hat{\rho}_{\boldsymbol{\lambda}}$ and $\tilde{\rho}_{\boldsymbol{\lambda}}$ as legitimate density matrices, adding proper normalization factors (e.g. introducing $p_{\boldsymbol{\lambda}}$ in Eq. \eqref{eq:decomp1}).
In the present case, we can simply write $\hat{H} = \sum_{\boldsymbol{\lambda}}  \tilde{H}_{\boldsymbol{\lambda}} \otimes \mathds{I}_{\boldsymbol{\lambda}} $, where the definition of $\hat{H}_{\boldsymbol{\lambda}}$ can be obtained replacing $\hat{\rho}$ with $\hat{H}$ in Eq. \eqref{eq:deftilted} and removing the factor $\dim \boldsymbol{\lambda}$.
We can use the decomposition above in the free energy \eqref{eq:partition}. Since the different $\boldsymbol{\lambda}$ subspaces are orthogonal, we have $ e^{- \beta \hat{H}} =  \sum_{\boldsymbol{\lambda}} e^{- \beta \tilde{H}_{\boldsymbol{\lambda}} \otimes \mathds{I}_{\boldsymbol{\lambda}} }$, which leads to
\begin{equation} \label{eq:partfunc}
    F = - \frac{1}{\beta} \ln \big\{  \sum_{\boldsymbol{\lambda}} \dim \boldsymbol{\lambda} 
    \Tr\big[e^{- \beta \tilde{H}_{\boldsymbol{\lambda}}}\big] \big\}\,,
\end{equation}
where $\dim \boldsymbol{\lambda}$ emerges from the trace of $\mathds{I}_{\boldsymbol{\lambda}}$.
We now consider $G= \text{S}_N$. To study the 
contributions of $\tilde{H}$ to Eq. \eqref{eq:partfunc} in the large $N$ limit we introduce the intensive free energy of the coarse-grained blocks
\begin{equation} \label{eq:energetic}
e(\boldsymbol{x}) \equiv - \lim_{N \rightarrow \infty} \frac{1}{\beta N} \ln\big\{\Tr[e^{- \beta \tilde{H}(\boldsymbol{x})}]\big\} \,,
\end{equation}
where $\tilde{H}(\boldsymbol{x}) = 
\tilde{H}_{\boldsymbol{\lambda}/N}$ is an operator acting on the space spanned by $\boldsymbol{\mu},h$. 
However, as discussed in Eqs. \eqref{eq:bound1}, \eqref{eq:limit1}, the intensive entropy associated with such contributions vanishes when $N\rightarrow \infty$, and $e(\boldsymbol{x})$ reduces to the intensive energy of the coarse-grained blocks.
Moreover, assuming that at least the ground state energy, $E_{\boldsymbol{x}}^0$, of $\tilde{H}(\boldsymbol{x})$ is extensive, 
$e(\boldsymbol{x})$ will be dominated by the ground state energy in every coarse-grained block:
\begin{align} \label{eq:rateerg}
e(\boldsymbol{x}) = \lim_{N \to \infty} \frac{E_{\boldsymbol{x}}^0}{N}.
\end{align}
This reduces the calculation of $e(\boldsymbol{x})$ to a much simpler ground state problem.
Using Eq. \eqref{eq:partfunc} with Eqs. \eqref{eq:rateent}, \eqref{eq:energetic}, we can express the intensive free energy as
\begin{equation} \label{eq:largedev}
   \lim_{N \to \infty} \frac{F}{N} = - \lim_{N \to \infty} \frac{1}{\beta N} \ln \int  e^{- N f(\boldsymbol{x})}  d\boldsymbol{x} = f(\boldsymbol{x}^*) \, ,
\end{equation}
where we introduced the free energy function
\begin{equation} \label{eq:IntFreeEnergy}
f(\boldsymbol{x}) \equiv e(\boldsymbol{x}) - \beta^{-1} s(\boldsymbol{x}) \;,
\end{equation}
and its minimum $\boldsymbol{x^*}=\operatorname*{arg\,min}_{\boldsymbol{x}} f(\boldsymbol{x})$.
Since the equilibrium density matrix of the system is $\hat{\rho} = e^{- \beta (\hat{H} - F)}$, using Eq. \eqref{eq:decomp1}), we find that
$p_{\boldsymbol{\lambda}}= \dim _{\boldsymbol{\lambda}} \Tr \left[e^{- \beta (\tilde{H}_{\lambda}-F)} \right]$.
This implies that 
\begin{align}
\lim_{N \to \infty} \frac{1}{N} \ln p_{\boldsymbol{\lambda}}= -\beta f(\boldsymbol{x}) \;,
\end{align}
which in turn implies that the entropy of the system can be calculated, using \eqref{VNentropy}, as 
\begin{equation} \label{eq:IntEntrop}
\lim_{N \to \infty} \frac{S}{N} = s(\boldsymbol{x^*}) \,,
\end{equation}
and the energy, $E=F+\beta^{-1} S$, as 
\begin{equation} \label{eq:IntEner} 
\lim_{N \to \infty} \frac{E}{N} 
= \lim_{N \to \infty} \frac{\Tr \left[ \hat{H} e^{- \beta (\hat{H}-F)}  \right]}{N} 
= e(\boldsymbol{x^*})  \,.
\end{equation}
The macroscopic thermodynamics of the system is thus fully characterized by the minimum of the free energy function \eqref{eq:IntFreeEnergy} that can be obtained from the knowledge of the ground state of the coarse-grained Hamiltonians $\tilde{H}(\boldsymbol{x})$ (see Eq.~\eqref{eq:rateerg}) and the universal $s(\boldsymbol{x})$ resulting from the structure of the permutation group (see Eq.~\eqref{eq:rateent}).

{\it Application.---}To illustrate the power of our approach, we use it to describe the phase diagram of the transverse-field Curie-Weiss model, consisting of $N$ spin-$1/2$ particles interacting via the Hamiltonian
\begin{equation} \label{eq:hamcw}
\hat{H} = - \omega \sum_{i=1}^N \hat{S}_z^{(i)} 
- \frac{\alpha}{N} \sum_{i,j=1}^N \hat{S}_x^{(i)} \hat{S}_x^{(j)},
\end{equation}
where $\hbar=1$, $\hat{S}_{z,x}^{(i)}$ are spin-1/2 operators, $\omega$ is the transverse magnetic field, and $\alpha$ is the Ising-type ferromagnetic (antiferromagnetic) interaction for $\alpha>0$ ($\alpha <0$). Despite its apparent simplicity, this model exhibits a highly nontrivial behavior resulting from the interplay of thermal and quantum fluctuations~\cite{chayes2008phase, jorg2010energy}. The model is permutation invariant, with $d=2$ in Eq. \eqref{eq:hilbert}.
In this case, $\boldsymbol{\lambda}$ is a $2$-dimensional partition of $N$, and can be fully parametrized in terms of a single rescaled variable $l = \frac{\lambda_1 - \lambda_2}{2 N} = \frac{\lambda_1}{N} - \frac{1}{2} \in [0,\frac{1}{2}]$. Physically, the representations $\boldsymbol{\lambda}$ are the subspaces corresponding to a definite eigenvalue $L$ of the total angular momentum operator $\hat{L}^{ 2} = \hat{L}^{ 2}_x + \hat{L}^{ 2}_y + \hat{L}^{ 2}_z$, where $\hat{L}_{x,y,z} = \sum_{i=1}^N \hat{S}^{(i)}_{x,y,z}$. The variable $l=L/N$ is then the rescaled total angular momentum. In particular, in the magnetic system it corresponds to the rescaled total magnetization.

For the model considered, the entropic factor $s(l)$ can be expressed using Eq.~\eqref{eq:rateent} as
\begin{align} \label{eq:rateentcw}
s(l)=-\left(\tfrac{1}{2}-l \right) \ln \left(\tfrac{1}{2}-l \right)-\left(\tfrac{1}{2}+l \right) \ln \left(\tfrac{1}{2}+l \right).
\end{align}
To compute the energetic factor $e(l)$ we use the mentioned correspondence between $l$ and the total angular momentum. The Hamiltonian~\eqref{eq:hamcw} can then be written in a block-diagonal form [see discussion below ~\eqref{eq:partition}] with 
\begin{equation}
\label{eq:reduced}
\tilde{H}_l = - \omega \hat{L}_z^{(l)} 
-\frac{\alpha}{N}  \hat{L}_x^{(l)} \hat{L}_x^{(l)}\,,
\end{equation}
where $\hat{L}^{(l)}_{x,z}$ are spin-$(lN)$ operators.
The matrix elements of \eqref{eq:reduced} in the $\hat{L}^{(l)}_{z}$ eigenbasis can be written in terms of the Clebsch-Gordan coefficients. After this step, for finite system sizes, Eqs.~\eqref{eq:energetic} and~\eqref{eq:rateerg} can be evaluated numerically. For $N\rightarrow \infty$, we can further use a quantum-classical correspondence for large $L=lN$ to calculate the ground state energy $E_l^0$ as the minimum energy of a corresponding classical Hamiltonian (see Appendix \ref{app:derenergeticcwan}). This yields
\begin{align} \label{eq:energeticcwan}
	e(l)=	\begin{cases}
			-\omega l & \text{for} \quad l \leq \frac{\omega}{2 \alpha}\,, \\
			-l^2 \alpha-\frac{\omega^2}{2 \alpha} & \text{for} \quad l>\frac{\omega}{2 \alpha}\,.
		\end{cases}
  \end{align}

\begin{figure}
    \centering
    \includegraphics[width=\columnwidth]{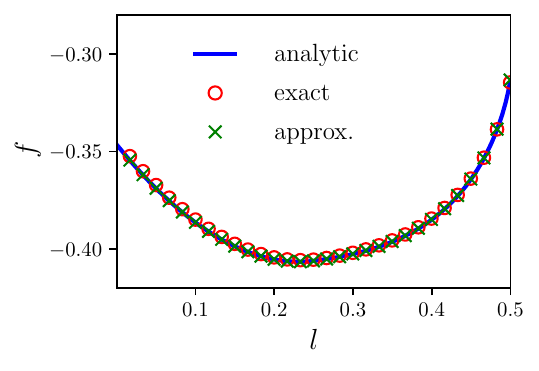}
    \caption{Free energy potential $f(l)$ calculated using analytic theory (blue solid line), exact numerical calculations (red circles), and approximate numerics (green crosses); see details in the main text. Parameters: $\alpha=1$, $\omega=0.5$, $\beta=2$, $N=1500$.}
    \label{fig:IF}
\end{figure}

To test our method, in Fig.~\ref{fig:IF} we present the free energy potential $f(l)$ calculated for large $N=1500$ using three different approaches: without any approximation using \eqref{eq:partfunc}, then with asymptotic formulas~\eqref{eq:rateent} and~\eqref{eq:rateerg}, and finally using the analytic expression~\eqref{eq:energeticcwan} for $e(l)$.  
All three approaches match well, which confirms the validity of our theory.

\begin{figure}
    \centering
\includegraphics[width= \columnwidth]{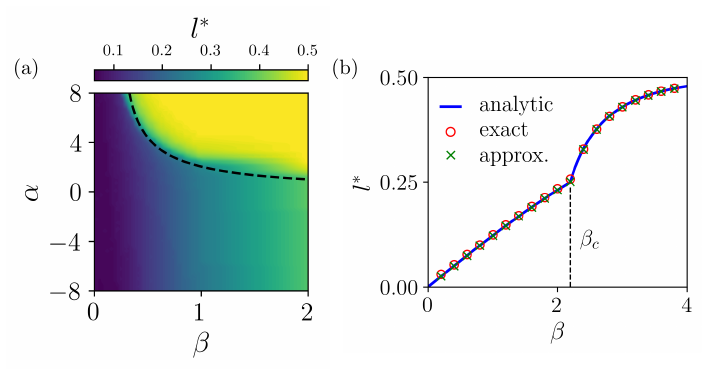}
    \caption{(a): Density plot $l^*$ in $(\beta,\alpha)$ coordinates, representing the phase diagram of the quantum Curie-Weiss model. Dashed line denotes the border of the ordered phase (top right corner of the figure). Plot obtained using exact numerics for $N=100$. (b): Cross section of the density plot for $\alpha=1$ and $N=1500$. Methods and symbols used as in Fig.~\ref{fig:IF}.
    }
    \label{fig:phase}
\end{figure}

In Fig.~\ref{fig:phase}~(a) we present the phase diagram of the model in $(\beta,\alpha)$ coordinates, specifically, the behavior of the argument $l^*$ minimizing the free energy potential, which corresponds to equilibrium total magnetization. As one can observe, $l^*$ is strongly enhanced in the ferromagnetic regime ($\alpha>0$) for $\beta>\beta_c$, where $\beta_c=2 \tanh^{-1}(\omega/\alpha)/\omega$ is the critical inverse temperature (top right corner of the plot). This region corresponds to the ordered phase, where the Ising interaction induces magnetization in the $x$ direction. Below $\beta_c$ and in the antiferromagnetic regime ($\alpha<0$), $l^*$ is well described by the formula for the magnetic-field-induced magnetization in the $z$ direction in the paramagnetic phase: $l^*=\tanh(\omega \beta/2)/2$. In Fig.~\ref{fig:phase}~(b) we plot a cross section of this graph for the constant $\alpha$, calculated using the same approaches as for Fig.~\ref{fig:IF}. As shown, the analytic solution exhibits a nonanalytic behavior at the phase transition point $\beta=\beta_c$, in agreement with both exact and asymptotic numerical results. 

{\it Final remarks.---}Our findings are particularly relevant to charaterize emergent phenomena in many-body quantum systems, where they can be used to drastically simplify problems that have so far resisted analytical or numerical treatment. We focused here on the permutation group, with an emphasis on the case $d=2$, but the scalings obtained can be straightforwardly applied to the case of all-to-all interacting $d$ level systems with $d \geq 2$. The case $d=3$ was recently experimentally studied using Rydberg atoms \cite{wild2023quantum}.
If $d$ is not finite, bounds of the form \eqref{eq:bound1}, \eqref{eq:bound2} still exist (see e.g. the Hardy-Ramanujan formula for counting the number of unrestricted partitions), and our theory can be easily extended to the case where the local Hilbert spaces have an extensive spectrum. It would also be interesting to study other groups, in particular those where $\ln\dim\boldsymbol{\lambda}$ is extensive; in such cases, we anticipate non-trivial phase transitions to occur. Finally, extending
our framework to open quantum systems could also be insightful for the perspective of building a quantum counterpart to classical macroscopic fluctuation theory \cite{derrida2007,bertini2015mft,bernard2021,Herpich2020,falasco2023macroscopic, meibohm2024minimum}.
\begin{acknowledgments}
This research was supported by the Luxembourg National Research Fund: A.S. by Project ThermoQO (C21/MS/15713841) and T.A. by ChemComplex (C21/MS/16356329).   
\end{acknowledgments}

\bibliographystyle{apsrev4-1}
\bibliography{bibliography}
\clearpage
\appendix

\section{Entropy and block structure of representations}

To derive Eq. \eqref{eq:entropy} we start by combining the decompositons in Eqs. 
\eqref{eq:decomp1}, \eqref{eq:decomp2}
\begin{equation}
\hat{\rho} = \sum_{\boldsymbol{\lambda}} \frac{p_{\boldsymbol{\lambda}}}{\dim \boldsymbol{\lambda}} \mathds{I}_{\boldsymbol{\lambda}} \otimes \tilde{\rho}_{\boldsymbol{\lambda}}.
\end{equation}
Inserting the equation above inside the definition of von Neumann entropy, we find
\begin{equation} \notag
S(\hat{\rho}) = - \Tr \big[  
\sum_{\boldsymbol{\lambda}} \frac{p_{\boldsymbol{\lambda}}}{\dim \boldsymbol{\lambda}} \mathds{I}_{\boldsymbol{\lambda}} \otimes \tilde{\rho}_{\boldsymbol{\lambda}} \ln \big(\sum_{\boldsymbol{\lambda}} \frac{p_{\boldsymbol{\lambda}}}{\dim \boldsymbol{\lambda}} \mathds{I}_{\boldsymbol{\lambda}} \otimes \tilde{\rho}_{\boldsymbol{\lambda}} \big) \big].
\end{equation}
Since $\tilde{\rho}_{\boldsymbol{\lambda}} \otimes \mathds{I}_{\boldsymbol{\lambda}}$ and $\tilde{\rho}_{\boldsymbol{\lambda}'} \otimes \mathds{I}_{\boldsymbol{\lambda}'}$ live in different blocks for $\boldsymbol{\lambda} \neq \boldsymbol{\lambda}' $, we can split the sum as
\begin{equation} \notag
S(\hat{\rho}) = - \sum_{\boldsymbol{\lambda}}\Tr \big[ \big(
 \frac{p_{\boldsymbol{\lambda}}}{\dim \boldsymbol{\lambda}} \mathds{I}_{\boldsymbol{\lambda}} \otimes \tilde{\rho}_{\boldsymbol{\lambda}}\big) \ln \big(\frac{p_{\boldsymbol{\lambda}}}{\dim \boldsymbol{\lambda}} \mathds{I}_{\boldsymbol{\lambda}} \otimes \tilde{\rho}_{\boldsymbol{\lambda}} \big) \big].
\end{equation}
We proceed by isolating the contributions of $\dim \boldsymbol{\lambda}$ and $p_{\boldsymbol{\lambda}}$ in the logarithm,
\begin{align} \label{eq:sumlog}
 &\Tr \big[ \big(
 \frac{p_{\boldsymbol{\lambda}}}{\dim \boldsymbol{\lambda}} \mathds{I}_{\boldsymbol{\lambda}} \otimes \tilde{\rho}_{\boldsymbol{\lambda}}\big) \ln \big(\frac{p_{\boldsymbol{\lambda}}}{\dim \boldsymbol{\lambda}} \mathds{I}_{\boldsymbol{\lambda}} \otimes \tilde{\rho}_{\boldsymbol{\lambda}} \big) \big]\\
 &=\Tr \big[ \big(
 \frac{\mathds{I}_{\boldsymbol{\lambda}} }{\dim \boldsymbol{\lambda}} \otimes \tilde{\rho}_{\boldsymbol{\lambda}}\big)\big]p_{\boldsymbol{\lambda}}\ln p_{\boldsymbol{\lambda}} \nonumber\\
 &-\Tr \big[\big(
 \frac{p_{\boldsymbol{\lambda}}}{\dim \boldsymbol{\lambda}} \mathds{I}_{\boldsymbol{\lambda}} \otimes \tilde{\rho}_{\boldsymbol{\lambda}}\big) \big]\ln \dim\boldsymbol{\lambda} \nonumber\\
 &+\Tr \big[
 \frac{p_{\boldsymbol{\lambda}}}{\dim \boldsymbol{\lambda}} \mathds{I}_{\boldsymbol{\lambda}} \otimes \tilde{\rho}_{\boldsymbol{\lambda}}\ln \tilde{\rho}_{\boldsymbol{\lambda}} \big] \, .\nonumber
\end{align}
Using that $\Tr[\mathds{I}_{\boldsymbol{\lambda}} \otimes \tilde{\rho}_{\boldsymbol{\lambda}}] = \dim \boldsymbol{\lambda}$, the Shannon entropy contribution $H(p)$ corresponds to the first term of the r.h.s. of Eq. \eqref{eq:sumlog} in the equation for $S(\hat{\rho})$.
The second term gives rise to the term proportional to $\ln\dim \boldsymbol{\lambda}$ in Eq. \eqref{eq:entropy}, while the remainder is the contribution proportional to $S(\tilde{\rho}_{\boldsymbol{\lambda}})$, using the property of the trace : $\Tr[A\otimes B]=\Tr[A]\Tr[B]$.

\label{app:entropy}
\section{What we can learn about $\mathcal{H}$ using representation theory}
\label{app:modules}

The irreducible representations of the permutation group have been found and described extensively \cite{sagan}.
This has been done in terms of {\it Young tableaux} that, at first glance, are not connected to the representations of $\text{S}_N$ in the physical Hilbert space \eqref{eq:hilbert}.
In this section we introduce Young tableaux and we study their connection with the physical space $\mathcal{H}$.
\subsection{Partitions, tableaux and tabloids}
Partitions are a crucial concept in the theory of the irreducible representations of $\text{S}_N$.
Given a number $N$, a partition $\boldsymbol{\lambda}$ is a weakly decreasing succession of integers that sums to $N$. For example, all the possible partitions for $N=4$ are listed below
\begin{align}
(1,1,1,1) \,, (2,1,1) \,, (2,2) \,, (3,1) \,, (4).
\end{align}
In contrast, $(1,1,2)$ does not represent a partition of $4$.
Given a partition $\boldsymbol{\lambda} = (\lambda_1,..\lambda_d)$
we can create Young tableaux as follows:

\begin{enumerate}
    \item The partition defines a \textit{shape}: for every $i=1,..,d$, create a row of $\lambda_i$ empty boxes.
The rows are conventionally ordered from up to down according to their index $i$ \footnote{This is the so-called ``english notation'' for Young tableaux.} and left-justified (that is, all the rows start from the same column on the left).
    \item The Young tableaux are obtained by filling the boxes using all the numbers $j$ between $1$ and $N$. 
\end{enumerate}

The partition $\boldsymbol{\lambda}$ is then called the shape of the tableaux created in this way.
Let's look at the example $N=3$. There are three partitions for $N=3$: $(1,1,1), (2,1), (3)$, corresonding respectively to the shapes
\begin{equation}
    \ytableaushort [] {\cdot,\cdot,\cdot}  \quad , \quad \ytableaushort [] {\cdot\cdot,\cdot} \quad , \quad \ytableaushort [] {\cdot\cdot\cdot}\label{1tableau} \, .
\end{equation}
For each of these shapes, there are 6 ways to fill them; here are a few examples:
\begin{equation}
    \ytableaushort [] {1,3,2} \quad , \quad \ytableaushort [] {3 1,2} \quad , \quad \ytableaushort [] {213} \label{1tableau2.0}
\end{equation}
The action of $\text{S}_N$ on Young tableaux is now defined in the following way: given a permutation $\sigma\in\text{S}_N$, the action of $\sigma$ on a Young tableau gives a Young tableau of identical shape in which the numbers $j$ have been replaced by their image $\sigma(j)$.
For example, consider the permutation $\pi$ defined as
$\pi(j) = j+1$ for $j < N$ and $\pi(N) = 1$.
For $N=3$, we can apply $\pi$ to the second tableau in Eq. \eqref{1tableau2.0} and obtain
\begin{equation}
  \pi\,\, \ytableausetup{boxsize=2em}
\ytableaushort [] { 3 1,2}\quad = \quad \ytableaushort [] {{\pi(3)} {\pi(1)},{\pi(2)}} \quad = \quad \ytableaushort [] {1 2,3}  \, .
\end{equation}

Now that we have introduced the Young tableaux, we need to explain two crucial points, that is, how to build a Hilbert space using them and how to define an action of $\text{S}_N$ in this space. After these two steps are clear, we will establish a link between this Young tableau defined space and our physical space $\mathcal{H}$ equipped with the physically motivated action of $\text{S}_N$ given by Eq. \eqref{eq:perm-states}.
Establishing this connection will allow us to justify the decomposition \eqref{eq:module2}.

A vector space based on tableaux can be obtained by associating every tableau to a new object, the {\it tabloid}.
Given a tableau $t$, the corresponding tabloid is the set of all tableaux that can be obtained from $t$ by permuting the elements in the same rows. Usually, the tabloid associated with $t$ is denoted $\{ t \}$.
For example, the tabloid associated to the tableau
\begin{equation}
    t=\ytableausetup{boxsize=normal,notabloids}
\ytableausetup{aligntableaux=center} \ytableaushort [] {12,3} 
\label{eq:ex-tableau}
\end{equation}
is
\begin{equation} \{ t \} \equiv
\ytableausetup{boxsize=normal,tabloids}
\ytableausetup{aligntableaux=center} \ytableaushort [] {12,3} \equiv \Big\{ \, \ytableausetup{boxsize=normal,notabloids}
\ytableausetup{aligntableaux=center} \ytableaushort [] {12,3} \,,\,\ytableausetup{boxsize=normal,notabloids}
\ytableausetup{aligntableaux=center} \ytableaushort [] {21,3} \, \Big\} ,  \end{equation}
where we also introduced, in the second member of the equation above, a new diagrammatic representation to represent a tabloid: it is simply obtained by removing all the vertical lines from the corresponding tableau. This representation illustrates the idea that tabloids are obtained by permuting the elements within each row of a tableau, so the position of a given element inside its row is not important. Consequently, different tableaux may yield the same tabloid.
Let's see another example, by choosing a slightly more complicated tableau
\begin{equation} 
    t=\ytableausetup{boxsize=normal,notabloids}
\ytableausetup{aligntableaux=center} \ytableaushort [] {13,25,4} \, .
\label{eq:ex-tableau2}
\end{equation}
The associated tabloid is
\begin{align} \notag \{ t \} =
\ytableausetup{boxsize=normal,tabloids}
\ytableausetup{aligntableaux=center} \ytableaushort [] {13,25,4} = \Big\{ \, \ytableausetup{boxsize=normal,notabloids}
\ytableausetup{aligntableaux=center} \ytableaushort [] {13,25,4} \,,\,\ytableausetup{boxsize=normal,notabloids}
\ytableausetup{aligntableaux=center} \ytableaushort [] {31,52,4} ,\,\ytableausetup{boxsize=normal,notabloids}
\ytableausetup{aligntableaux=center} \ytableaushort [] {31,25,4},\,\ytableausetup{boxsize=normal,notabloids}
\ytableausetup{aligntableaux=center} \ytableaushort [] {13,52,4} \, \Big\} .  \end{align}
The tabloids can now be used to define a vector space, as anticipated above. Consider a partition $\boldsymbol{\mu}$ of $N$ (we use a different notation than $\boldsymbol{\lambda}$ for reasons which will be clear later). A {\it permutation module} $M'_{\boldsymbol{\mu}}$ is by definition the vector space where the basis elements are all the tabloids associated to the partition $\boldsymbol{\mu}$, that is, all the tabloids which can be obtained from tableaux of shape $\boldsymbol{\mu}$.
Notice that, even if this object may seem different from the $M_{\boldsymbol{\mu}}$ introduced in the main text, we will soon show that the two spaces are isomorphic (see next subsection).
Before proceeding further, let us see an example of a permutation module. Choose a partition $\boldsymbol{\mu} = (2,1)$, we have $6$ possible tableaux and $3$ possible tabloids, which are
\begin{equation}
  \ytableausetup{boxsize=normal,tabloids}
\ytableausetup{aligntableaux=center} \ytableaushort [] {12,3}  \, , \,   \ytableausetup{boxsize=normal,tabloids}
\ytableausetup{aligntableaux=center} \ytableaushort [] {13,2}   \, , \, 
  \ytableausetup{boxsize=normal,tabloids}
\ytableausetup{aligntableaux=center} \ytableaushort [] {23,1}  \, .
\end{equation}
Then, every vector $w$ in the permutation module $M_{(2,1)}$ can be written as a linear combination of those three tabloids,
\begin{equation}
w = a_1  \, \ytableausetup{boxsize=normal,tabloids}
\ytableausetup{aligntableaux=center} \ytableaushort [] {12,3}  + a_2 \, \ytableausetup{boxsize=normal,tabloids}
\ytableausetup{aligntableaux=center} \ytableaushort [] {13,2}  + a_3 \,
  \ytableausetup{boxsize=normal,tabloids}
\ytableausetup{aligntableaux=center} \ytableaushort [] {23,1} \, , \label{example1}
\end{equation}
where $a_1,a_2,a_3\in\mathbb{C}$. 
We can now define an action of our permutation group on $M'_{\boldsymbol{\mu}}$.
We have already discussed how a permutation acts on tableaux; to compute the permutation of a tabloid we compute the permutation of the original tableaux and build its tabloid:
\begin{equation}
    \pi \{t \} = \{ \pi t \}.
\end{equation}
This naturally defines an action over the vector spaces $M'_{\boldsymbol{\mu}}$, for instance the permutation $\pi_{12}$ exchanging the numbers $1$ and $2$ applied to \eqref{example1} gives
\begin{align} \notag
\pi_{12} w & = a_1 \pi_{12} \, \ytableausetup{boxsize=normal,tabloids}
\ytableausetup{aligntableaux=center} \ytableaushort [] {12,3}  + a_2  \pi_{12} \, \ytableausetup{boxsize=normal,tabloids}
\ytableausetup{aligntableaux=center} \ytableaushort [] {13,2}  + a_3 \pi_{12} \,
  \ytableausetup{boxsize=normal,tabloids}
\ytableausetup{aligntableaux=center} \ytableaushort [] {23,1} \,  \\
& = a_1  \, \ytableausetup{boxsize=normal,tabloids}
\ytableausetup{aligntableaux=center} \ytableaushort [] {12,3}  + a_2   \, \ytableausetup{boxsize=normal,tabloids}
\ytableausetup{aligntableaux=center} \ytableaushort [] {23,1}  + a_3  \,
  \ytableausetup{boxsize=normal,tabloids}
\ytableausetup{aligntableaux=center} \ytableaushort [] {13,2} \, . \label{example1__}  
\end{align}
We have so defined, for every partition $\boldsymbol{\mu}$ of $N$, a vector space $M'_{\boldsymbol{\mu}}$ in which the action of $\text{S}_N$ is well defined.
We can then apply group representation theory in order to decompose each permutation module in irreducible representations of $\text{S}_N$ \cite{sagan}. 
The only thing left to show is that the permutation modules $M'_{\mu}$ are isomorphic to the $M_{\mu}$ introduced in the main text, meaning that there exists a bijection between those two objects which preserves the action of the group.
For a more in-depth read about the topics of this section, see pages 53-57 of \cite{sagan}.

\subsection{Isomorphism between permutation modules and subspaces with fixed occupation numbers}
\label{subsec:isom}

In this section, we build a mapping between the abstract permutation modules $M'_{\boldsymbol{\mu}}$ and the physical subspaces  $M_{\boldsymbol{\mu}}$ introduced in Eq. \eqref{eq:hilbert}.  

Consider a subspace $M_{\boldsymbol{\mu}}$ of $\mathcal{H}_N^{(d)}$, consisting of states with fixed occupation numbers given by $\boldsymbol{\mu}=(\mu_1,...,\mu_d)$. 
Without loss of generality, we can assume that the occupation numbers are ordered in a decreasing fashion.
The first thing to prove if we want to show that  $M_{\boldsymbol{\mu}}$ is isomorphic to $M'_{\boldsymbol{\mu}}$, is that the dimensions of the two spaces are the same.
The dimension of the space with occupation numbers $\boldsymbol{\mu}$ is given by the well known combinatorial formula
\begin{equation} \label{eq:dimmu}
    \dim M_{\boldsymbol{\mu}} = \frac{N!}{\mu_1!... \mu_N!}.
\end{equation}
To compute the dimension of $M'_{\boldsymbol{\mu}}$, recall that its basis elements are given by all possible tabloids associated to the partition $\boldsymbol{\mu}$, so it is sufficient to count the number of such tabloids.
By construction, a tabloid is fully characterized by assigning each number $1,..,N$ of its filling to the row $1,.., d$ where this number belongs. 
The number of ways in which this assignment can be done coincides with the number of ways $N$ particles can be placed in $d$ boxes, that is equal to the r.h.s of \eqref{eq:dimmu}.

From this argument, a natural connection between the elements of $M_{\boldsymbol{\mu}}$ and $M'_{\boldsymbol{\mu}}$, appears to be the one associating the levels of the single-particle Hilbert spaces with the rows of the tabloids. To see this, let us introduce a mapping $\theta$ between the product basis vectors of $\mathcal{H}$ and the tabloids as follows. Given a basis vector $\ket{e_1} \otimes \ket{e_2} ... \otimes \ket{e_n}$, where we recall that $|e_i\rangle$ is a basis vector for $\mathbb{C}_d^{(i)}$, with $e_i\in\{1,...,d\}$, let us map it to

\begin{equation} \label{eq:defisom}
\theta (  \ket{e_1} \otimes \ket{e_2} ... \otimes \ket{e_n} ) =  
\ytableausetup{boxsize=normal,tabloids} \ytableausetup{aligntableaux=center} {\ytableausetup{centertableaux}
\, \begin{ytableau}
i \,& s.t. & &  e_i =1 & \\
i \,& s.t. & &  e_i =2 & \\
& .... & .... & \\
i \,& s.t. & &  e_i =d &
\end{ytableau}},
\end{equation}
meaning that the indices of the particles in the level $1$ are placed in the first row of the tabloid, the indices of the particles in level $2$ are placed in the second row, and so on.
Here are some examples of the mapping for vectors with $\boldsymbol{\mu} = (2,2,1)$:

\begin{align}
\theta ( \ket{12123} ) = 
\ytableausetup{boxsize=normal,tabloids} \ytableausetup{aligntableaux=center} \ytableaushort [] {13,24,5}, \quad \theta ( \ket{13221} ) =  \ytableausetup{boxsize=normal,tabloids} \ytableausetup{aligntableaux=center} \ytableaushort [] {15,34,2} ,
\end{align}
where we dropped the $\otimes$ symbols to alleviate the notation.
The mapping $\theta$ is linear and sends vectors with occupation $\boldsymbol{\mu}$ in tabloids with shape $\boldsymbol{\mu}$, so we have
$\theta \big(M_{\boldsymbol{\mu}} \big) \subseteq M'_{\boldsymbol{\mu}}$.
In addition, it is easy to prove using \eqref{eq:defisom} that every standard basis vector of $M_{\boldsymbol{\mu}}$ is sent to a different tabloid in $M'_{\boldsymbol{\mu}}$, so $\theta$ is injective.
Since $M_{\boldsymbol{\mu}}$ and $M'_{\boldsymbol{\mu}}$ have the same dimension, we conclude that $\theta$ is an isomorphism of vector spaces.
It remains to prove that $M_{\boldsymbol{\mu}}$ and $M'_{\boldsymbol{\mu}}$ are not only the ``same'' vector space, but they behave in the same way under the action of $\text{S}_N$.
Formally, we have to prove that $\theta$ is a group homomorphism:
\begin{equation} \label{eq:grouphom}
\theta(\pi v) = \pi (\theta v) \, \, \, \forall \, v \in M_{\boldsymbol{\mu}}, \pi \in \text{S}_N.
\end{equation}
Since the transpositions (i.e. permutations of $2$ elements) generate all $\text{S}_N$ it is sufficient to prove Eq. \eqref{eq:grouphom} for the transpositions.
This means that exchanging two indices of $v \in M_{\boldsymbol{\mu}}$ and then applying $\theta$, should produce the same result as directly exchanging those indices in $\theta(v) \in M'_{\boldsymbol{\mu}}$.
This is immediate from the definition of $\theta$ in \eqref{eq:defisom}, and ends the proof.

We conclude that
the physical space with fixed occupation numbers $M_{\boldsymbol{\mu}}$ is isomorphic (in the sense of groups) to the permutation module $M'_{\boldsymbol{\mu}}$.
Since their structure under the action of the group $\text{S}_N$ is the same, they have the same decomposition in irreducible representations.
The decomposition of $M'_{\boldsymbol{\mu}}$ in irreducibles is given by a result from group representation theory known as {\it Young's rule} \cite{sagan}, therefore we can apply the same decomposition for $M_{\boldsymbol{\mu}}$ in the main text.

\section{Derivation of \eqref{eq:rateent} and \eqref{eq:limit1}}\label{app:semistand}

We begin by providing details on the derivation of \eqref{eq:rateent}. 
This derivation will make use of the concept of Young tableau introduced in section~\ref{app:modules} above and will make use of the {\it Hook length formula} \cite{fultonharris}.
Consider a partition $\boldsymbol{\lambda}=(\lambda_1,...,\lambda_d)$ of $N$. As explained in section~\ref{app:modules}, we may associate an empty Young tableau of $N$ to that partition. Let's denote with $(i,j)$ the coordinates of each box of that tableau, where $i,j$ correspond respectively to the line and column indices, with $(1,1)$ the coordinates of the top left box. To each box $(i,j)$, we associate a \textit{Hook length} $h_{\boldsymbol{\lambda}}(i,j)$, equal to the number of $(l,m)$ such that $l=i$ and $m\geq j$ or $a\geq i$ and $m=j$. For example, the Hook length of the box $(3,2)$ in the tableau below is equal to $6$, and represented by the black line:

$$ 
\ytableausetup{boxsize=normal,notabloids}
\begin{ytableau}
\cdot &  \cdot & \cdot & \cdot & \cdot & \cdot \\
\cdot &  \cdot & \cdot & \cdot & \cdot  \\
\cdot & \hskip 14mm  {\rule{12mm}{2pt}}\makebox[0pt][l]{\hspace{-13mm}\rule[-17.5mm]{2pt}{18mm}}  & \cdot & \cdot \\
\cdot & \cdot & \cdot & \cdot \\
\cdot & \cdot & \cdot \\
\cdot & \cdot
\end{ytableau} $$

The Hook length formula \cite{fultonharris} provides an expression for the dimension of the irreducible labelled by $\boldsymbol{\lambda}$,
\begin{equation}
    \mbox{dim}\boldsymbol{\lambda}=\frac{N!}{\prod_{(i,j)} h_{\boldsymbol{\lambda}}(i,j)} \, ,
    \label{eq:def-diml}
\end{equation}
where the product runs over all the boxes of the Young tableau of shape $\boldsymbol{\lambda}$. Taking the logarithm of \eqref{eq:def-diml}, using the Stirling formula $N!
=\sqrt{2\pi N}(N/e)^N(1+\mathcal{O}(1/N))$ and noticing that there are $N$ terms in the product in \eqref{eq:def-diml}, we may write
\begin{align}
\frac{\ln\mbox{dim}\boldsymbol{\lambda}}{N} = -1-\frac{1}{N}\sum_{(i,j)}\ln\left(\frac{h_{\boldsymbol{\lambda}}(i,j)}{N}\right)+\mathcal{O}\left(\frac{\ln N}{N}\right) \, . \label{eq:i-sigma-aux}
\end{align}
Let's now show that, after a proper rescaling, the function on the r.h.s. in \eqref{eq:i-sigma-aux} has a dominant contribution in the large $N$ limit that is constant in $N$.

We begin by writing explicitly the value of $h_{\boldsymbol{\lambda}}(i,j)$. The length of the $i-$th line of a tableau of shape $\boldsymbol{\lambda}=(\lambda_1,...,\lambda_d)$ is equal to $\lambda_i$. Let's note $\lambda_j^T$ the length of the $j-$th column. Then, we may write
\begin{equation}
    h_{\boldsymbol{\lambda}}(i,j)=\lambda_i-j+\lambda_j^T-i \, .
\end{equation}
Before going further, we point out that, in the case where $d=1$, $\dim \boldsymbol{\lambda}$ reaches its minimal value $\dim \boldsymbol{\lambda}=1$, and that for any $N>d>1$, $\dim \boldsymbol{\lambda}>1$. Let's assume further on that $N>d>1$. In this case, we may take the continuous limit of the sum in $j$ in the double sum in \eqref{eq:i-sigma-aux}: introducing the rescaled variable $u\equiv j/N$, and replacing, for every $i\in\{1,...,d\}$, 
\begin{equation}
    \frac{1}{N}\sum\limits_{j=1}^N \ln\left(\frac{h_{\boldsymbol{\lambda}}(i,j)}{N}\right)\rightarrow \int_0^{\lambda_i/N} du \ln\left(\frac{\lambda_i}{N}-u+\frac{\lambda_j^T-i }{N}\right) \, .
    \end{equation}
Since, for all $i,j$, $|\lambda_j^T-i|\leq d$, we may expand the logarithm,
\begin{equation}
    \ln\left(\frac{\lambda_i}{N}-u+\frac{\lambda_j^T-i }{N}\right) = \ln\left(\frac{\lambda_i}{N}-u\right) + \mathcal{O}\left(\frac{d}{N}\right) \, .
\end{equation}
Using the rescaled variable $\boldsymbol{x}=\boldsymbol{\lambda}/N$, with $x_i=\lambda_i/N$ being the elements of the rescaled partition $\boldsymbol{x}$, we may define the rate function
\begin{align} \nonumber
    s(\boldsymbol{x})\equiv &-1-\sum\limits_{i=1}^d\int\limits_0^{x_i} du\ln(x_i-u)= \\ &-\sum_{i=1}^d x_i \ln x_i ,
\end{align}
where we used $\sum_{i=1}^d x_i=1$. Consequently, 
%
\begin{equation}
    \dim (N\boldsymbol{x}) \asymp e^{Ns(\boldsymbol{x})} \, ,
    \label{eq:dim-lambda-rate}
\end{equation}
which is what we wanted to prove. 
%
We conclude this section with a comment on the connection between the expression \eqref{eq:dim-lambda-rate} and the Vershik-Kerov theorem in representation theory \cite{kerov1986}. 
The original theorem is stated under more general conditions than what we require here. Applied to our case, the theorem states that the renormalized lengths $x_i$ can be interpreted as random variables, which obey a law of large numbers: for any $\epsilon>0$ and all $j\in\{1,...,d\}$,
\begin{equation}
    \lim_{N\to+\infty} \mbox{dim}\{N\boldsymbol{x}:\Big|x_j-\frac{1}{d}\Big|\geq \epsilon\}/ \dim(N \boldsymbol{x}^*) =0 \, ,
\end{equation}
where $\mbox{dim}(N\boldsymbol{x}^*)$ is the maximal dimension possible, in our case achieved when all the $x_i$ are equal to $1/d$.

\medskip

We now turn to the bound \eqref{eq:limit1}, which uses properties of the Schur polynomials. Schur polynomials are homogeneous symmetric polynomials, of $d$ variables $(x_1,...,x_d)$, indexed by a partition $\boldsymbol{\lambda}=(\lambda_1,...,\lambda_d)$. 
The Schur polynomials form a linear basis of symmetric polynomials and play an important role in group representation theory. It can be shown \cite{fultonharris} that, given a partition $\boldsymbol{\lambda}=(\lambda_1,...,\lambda_d)$, the corresponding Schur polynomial evaluated at $(y_1,...,y_d)=(1,...,1)$ is equal to the number of irreducible representations labeled by $\boldsymbol{\lambda}$ that appear in the decomposition of a permutation module $M_{\boldsymbol{\mu}}$,
\begin{equation}
   P_{\boldsymbol{\lambda}}(1,...,1) = \sum\limits_{\boldsymbol{\mu}} K_{\boldsymbol{\lambda\mu}} ,
\end{equation}
where $K_{\boldsymbol{\lambda \mu}}$ is the multiplicity factor introduced in Eq. \eqref{eq:module2}.
Using the Weyl character formula, it can be shown that \cite{fultonharris}
\begin{equation} \vspace{0.2cm}
    P_{\boldsymbol{\lambda}}(1,...,1) = \prod_{1\leq i<j\leq d}\left(\frac{\lambda_i-\lambda_j}{j-i}+1\right) \, .
    \label{eq:schur1}
\end{equation}
Since $\sum_{j=1}^d \lambda_j=N$, and recalling that $\lambda_1\geq\lambda_2\geq...\geq \lambda_d$, we may bound
$
    \frac{\lambda_i-\lambda_j}{j-i} +1 \leq N \, , 
$
hence, since the product in \eqref{eq:schur1} contains $d(d-1)/2$ terms,
\begin{equation}
    P_{\boldsymbol{\lambda}}(1,...,1) \leq N^{d(d-1)/2} \, , 
\end{equation}
which leads to the bound \eqref{eq:limit1}.
\vspace{0.3cm}

\section{Derivation of~\eqref{eq:energeticcwan}}
\label{app:derenergeticcwan}
Using the principle of quantum-classical correspondence, for large values of the angular momentum $L=lN$, the Hamiltonian~\eqref{eq:reduced} can be replaced with its classical counterpart
\begin{equation}
\label{eq:reducedcl}
\tilde{H}_l^\text{cl} = - \omega L_z-\frac{\alpha}{N}  L_x^2,
\end{equation}
where $L_{x,y,z}$ are the classical angular momentum components obeying $L_{x,y,z} \in [-L,L]$ and $L^2=L_x^2+L_y^2+L_z^2$. The ground state energy $E_l^0$ corresponds to the minimum of~\eqref{eq:reducedcl} over $L_{x,y,z}$. Therefore, using Eq.~\eqref{eq:rateerg}, the energy term $e(l)$ can be calculated as
\begin{align}
e(l)= \min_{l_z \in [-l,l]} \left[-\omega l_z-\alpha \left(l^2-l_z^2 \right) \right],
\end{align}
where $l_{x,y,z}=L_{x,y,z}/N$. The solution yields Eq.~\eqref{eq:energeticcwan}.

\end{document}